# Modelling Microtubules in the Brain as
# *n*-qudit Quantum Hopfield Network and Beyond

*Dayal Pyari Srivastava[1], Vishal Sahni[2], Prem Saran Satsangi[3]*

**ABSTRACT**
The scientific approach to understand the nature of consciousness revolves around the study of human brain. Neurobiological studies that compare the nervous system of different species have accorded highest place to the humans on account of various factors that include a highly developed cortical area comprising of approximately 100 billion neurons, that are intrinsically connected to form a highly complex network. Quantum theories of consciousness are based on mathematical abstraction and Penrose-Hameroff Orch-OR Theory is one of the most promising ones. Inspired by Penrose-Hameroff Orch-OR Theory, Behrman et. al. (Behrman, 2006) have simulated a quantum Hopfield neural network with the structure of a microtubule. They have used an extremely simplified model of the tubulin dimers with each dimer represented simply as a qubit, a single quantum two-state system. The extension of this model to *n*-dimensional quantum states, or *n*-qudits presented in this work holds considerable promise for even higher mathematical abstraction in modelling consciousness systems.

## 1. Microtubules as Quantum Information / Computation Processing Devices : Penrose-Hameroff Orchestrated Objective Reduction (Orch. OR Theory)

Penrose and Hameroff (2011) describe microtubules as self-assembling polymers of the peanut-shaped (4 nm x 8 nm x 5 nm) protein dimer tubulin, each tubulin hetero dimer molecule (110,000 atomic mass units) being composed of an alpha and beta monomer, and is a polar molecule with its positive end near the $\beta$-subunit. Typically, thirteen linear tubulin chains ("protofilaments") align side-to-side to form hollow microtubule cylinders (25 nm diameter) with two types of hexagonal lattices A and B. The hollow microtubule cylinders of about 25 nm in diameter range from 200 nm to 25 micron in length. Structure of a microtubule and arrangement of its tubulins is given in Figure 1. Since there are at least 32 states, i.e., 5 bits (or 5/8 byte) of information per tubulin dumer molecule and 13 dimers (of 8 nm length) per each ring of microtubule-cylinder with 1250 rings per midsize (1-micron long) microtubule, the resulting information storage capacity is approximately 10 kilobytes per microtubule. Given 10,000 microtubules (or 100 million tubulin-dimers) per neuron, it represents 100 megabytes of processing power per neuron which translates into a total information storage capacity of human brain (with 100 billion neurons) of 10 exabytes (i.e. of the order of at least 1 exabyte = $10^{18}$ bytes).

---

[1] Department of Physics and Computer Science & Quantum-Nano Systems Centre, Dayalbagh Educational Institute, Dayalbagh, Agra, India
[2] Quantum-Nano Systems Centre & Centre for Consciousness Studies, Dayalbagh Educational Institute, Dayalbagh, Agra, India (email : vishalsahni.dei@gmail.com)
[3] Chairman, Advisory Committee on Education, Dayalbagh Educational Institutions, Dayalbagh, Agra, India and Member, Editorial Board, International Journal of General Systems since inception.

According to Penrose-Hameroff Orch. OR (Orchestrated Objective Reduction) theory, each tubulin molecule can exist as quantum superposition (i.e. quantum bit or qubit) of both states (black and white) coupled to London force dipole in hydrophobic pocket. Furthermore the A-lattice has multiple winding patterns which intersect on protofilaments at specific intervals matching the Fibonacci series found widely in nature and possessing a helical symmetry enabling topological quantum computing.

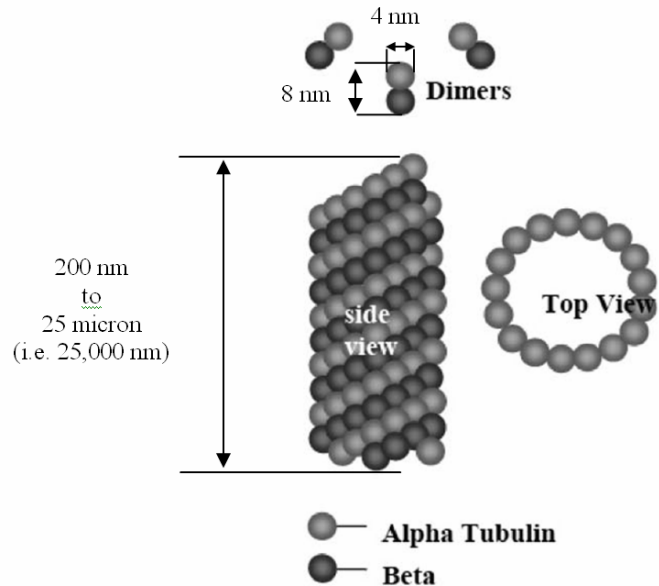

**Figure 1**    Structure of a Microtubule and the arrangement of its tubulins

Penrose and Hameroff (2011) estimate approximately $10^8$ tubulins in each neuron which switch and oscillate in the range of $10^7$ per second. This gives an information capacity as the single-cell value at the microtubule level of $10^{15}$ operations per second per neuron. The total brain capacity ($10^{11}$ neurons, $10^3$ synapses per neuron, $10^2$ transmissions per synapse per second) would thus potentially translate at the microtubule level as $10^{26}$ operations per second in comparison to the earlier estimates of AI community for the information processing capacity of the entire brain of $10^{16}$ operations per second at the level of neurons, synapses and their transmission-rate per second.

If each tubulin dimer functions as a quantum bit and not a classical bit processor, the computational power becomes almost unimaginably vast. It has been claimed that as few as 300 quantum bits (qubits), have the same computational power as a hypothetical classical computer comprised of as many processing units as there are particles in the universe (Steane and Rieffel, 2000).

Penrose-Hameroff Orchestrated Objective Reduction (Orch-OR) has three parts : the Gödel Part, the Gravity Part, and the Microtubule Part.

The Gödel Part is an effort to use the famous Gödel's Incompleteness Theorem [Any finite set of rules that encompass the rules of arithmetic is either inconsistent or incomplete; it entails either statements that can be proved to be both true and false, or statements that cannot be proved to be either true or false] to prove that human beings have intellectual powers of "non-computable" thought and understanding that they could not have if they functioned in accordance with the principles of classical physical theory. Proving this would reaffirm a conclusion of the von Neumann formulation of quantum theory, namely, that a conscious human being can behave in ways that a classical mechanical model cannot. [However, the Gödel Part cannot now be regarded as having been established successfully].

The Gravity Part addresses a key question pertaining to quantum dynamics : Exactly when do the sudden quantum jumps occur? The Diösi-Penrose (DP) expectation is that Objective Reduction (OR) occurs when the overall separation (the product of the temporal separation T with the spatial separation S) reaches a critical amount given by the Planck-Dirac constant $\hbar =$ Planck constant $h/2\pi$, such that the quantity $S$ is given by : $S \approx E_G$ (the gravitational self-energy) of the difference between the mass distributions of the two superposed state and $T = Z$ (the life-time) $= \hbar/E_G$. Penrose-Hameroff (PH) theory proposes that this time interval is the duration of time for which Nature will endure this bifurcation of space-time structure into the two, incompatible parts, before jumping to one or the other of these two forms. This conjectured rule is based on two very general features of Nature : Planck's universal constant of action $h$ and Newton-Einstein universal law of gravitation. It invokes quantum gravity attempting to combine quantum theory with Einstein's theory of gravity, namely General Relativity.

Moreover, according to Orchestrated Objective Reduction (Orch OR), this is accompanied by an element of proto-consciousness.

The best known temporal correlate for consciousness is gamma synchrony, EEG / MEG, 30-90 Hz, often referred to as coherent 40 Hz, representing a succession of 40 or so conscious moments per second ($\tau = 25$ milliseconds). Global macroscopic states such as superconductivity ensue from quantum coherence (with attendant superposition) among only very small fractions (say 1%) of components (six tubulins per neuron). For $\tau = 25$ msec, 20,000 such neurons would be required to elicit OR. In human brain, cognition and consciousness are, at any one time, thought to involve tens of thousands of neurons (10,000 to 1,00,000 neurons) which may be widely distributed throughout the brain. At 80 Hz or higher frequency, associated with Tibetan or other meditators, expanded awareness states of consciousness might be expected with more neuronal brain involvement.

Does this rule have any empirical support? An affirmative answer can be provided by the Microtubule Part of Penrose-Hameroff (PH) theory by linking Diosi-Penrose objective reduction rule to Hameroff's belief that consciousness is closely linked to the microtubular structure of the neuron.

## 2. Topological Quantum Computation

Topological quantum computing in 'Orch OR' is shown in Figure 2 (Penrose and Hameroff, 2011).

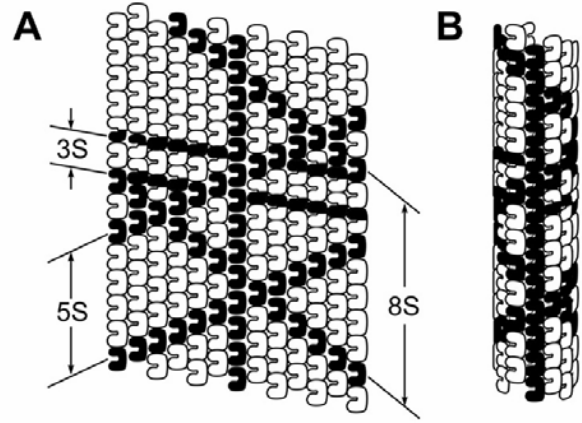

**Figure 2** extending microtubule A-lattice hydrophobic channels results in helical winding patterns matching Fibonacci geometry (Fibonacci series : e.g. 1, 1, 2, 3, 5, 8 etc. in which each Fibonacci number is the sum of the preceding numbers). Bandyopadhyay (2011) has evidence for ballistic conductance and quantum interference along such helical pathways which may be involved in topological quantum computing. Quantum electronic states of London forces in hydrophobic channels result in slight superposition separation of atomic nuclei, sufficient $E_G$ for Orch OR. This image may be taken to represent superposition of four possible topological qubits which, after time T = tau = $\frac{\hbar}{E_G}$, will undergo OR, and reduce to specific pathway(s) which then implement function. B-lattice microtubules have a vertical seam dislocation.

In the quantum theory, the quantum state of the *n* indistinguishable particles (e.g. quasiparticle, anyons) belongs to a Hilbert space that transforms as a unitary representation of the braid group $B_n$ of *n* strands (Sahni, Lakshminarayanan and Srivastava, 2011).

The braid group $B_n$ can be presented as a set of generators that obey particular defining relations. To understand the defining relations, we may imagine that the *n* (say 3) particles occupy *n* (say 3) ordered positions (labeled 1, 2, 3, . . ., *n*) arranged on a line. Let $\sigma_1$ denote a counterclockwise exchange of the particles that initially occupy positions 1 and 2, let $\sigma_2$ denote a counterclockwise exchange of the particles that initially occupy positions 2 and 3, and so on. Any braid can be constructed as a succession of exchanges of neighboring particles; hence $\sigma_1, \sigma_2, ..., \sigma_{n-1}$ are the group generators.

The second, slightly more subtle type of relation is

$$\sigma_j \sigma_{j+1} \sigma_j = \sigma_{j+1} \sigma_j \sigma_{j+1}, \qquad j = 1, 2, ..., n-2$$

which is called the Yang-Baxter relation. We can verify the Yang-Baxter relation by drawing the two braids $\sigma_1\sigma_2\sigma_1 = \sigma_2\sigma_1\sigma_2$ on a piece of paper (Fig. 3), and observing that both describe a process in which the particles initially in positions 1 and 3 are exchanged counterclockwise about the particle labeled 2, which stays fixed — i.e., these are topologically equivalent braids.

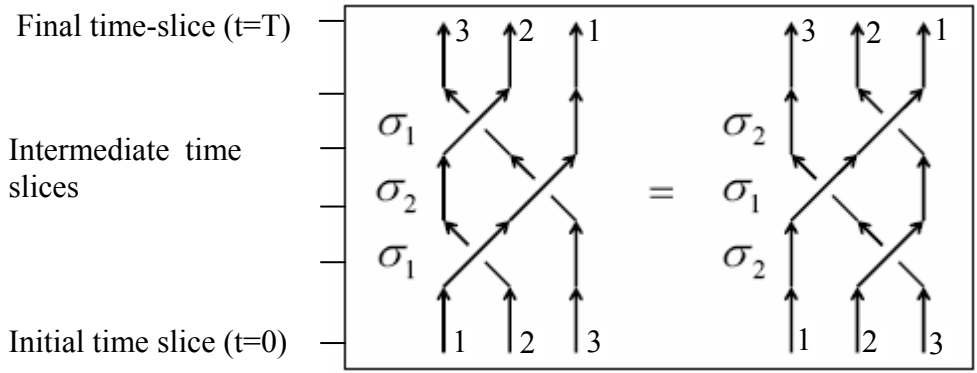

**Figure 3**   A simple diagram showing the two braids $\sigma_1\sigma_2\sigma_1 = \sigma_2\sigma_1\sigma_2$ with $n=3$ strands each

The most important issues facing topological quantum computation are twofold : (1) finding or identifying a suitable system with the appropriate topological properties, that is, non-abelian statistics, to enable quantum computation, and (2) figuring out a scheme to carry out the braiding operations necessary to achieve the required unitary transformations. The only topological system known to exist in nature is the quantum Hall regime. Thus, there is a strong need to find other systems satisfying the above two criteria for topological quantum computation.

Topological quantum computation can then be carried out by moving quasi particles around one another in two space dimensions. The quasi particle world-lines form topologically nontrivial braids in three (= 2 + 1) dimensional space-time, and because these braids are topologically robust (i.e., they cannot be unbraided without cutting one of the strands) the resulting computation is protected against error.

An Orch. OR qubit based on topological quantum computing specific to microtubule polymer geometry was suggested by Hameroff et. al. in 2002 (Penrose and Hameroff, 2011). Conductances along particular microtubule lattice geometry, e.g. Fibonacci helical pathways, were proposed to function as topological bits and qubits. Bandopadhyay (2011) has preliminary evidence for ballistic conductance along different, discrete helical pathways in single microtubules.

## 3. Modelling Mircotubules as Quantum Hopfield Networks

Hopfield neural networks are a class of neural network models where non-linear graded response neurons organized into networks with effectively symmetric synaptic connections are able to implement interesting algorithms, thereby introducing the concept of information storage in the stable states of dynamical systems. The dynamics of the state-space trajectory as well as time domain evolution of sensitivities of the states with respect to circuit parameters using system dynamics simulation has been extensively studied (Kumar and Satsangi, 1993).

Behrman et. al. (Behrman, 2006) have simulated a quantum Hopfield neural network with the structure of a microtubule. They have used an extremely simplified model of the tubulin dimers: each is represented simply as a qubit, a single quantum two-state system, the quantum analog of a classical bit. A qubit, unlike a bit, can exist in a superposition of the two states. They have included Coulombic interactions between qubits, and thermal effects for the qubits themselves. This is a kind of quantum Hopfield network. Each of the individual processing elements or neurons (here, the qubits) interacts with each of the others (a fully interconnected network); the elements (qubits) are initialized in some state, and allowed to evolve towards a local minimum. These networks store information and are also called "associative memory" or "content addressable memory" models. The final stable state is a pattern that is recalled by the network.

Inspired by Penrose-Hameroff Orch. OR theory, Behrman et. al. (Behrman, 2006) use an extremely simplified model of tubulin dimers, each represented simply as a qubit, a single quantum two-state system. In order to obtain the Quantum Hopfield Net (QHN), they consider an array of $N$ qubits. Hamiltonian or energy function operator for each qubit $j$ is given by

$$H_j = K\sigma_x + A\sigma_z \qquad (1)$$

where $\sigma_x = \begin{bmatrix} 0 & 1 \\ 1 & 0 \end{bmatrix}$ and $\sigma_z = \begin{bmatrix} 1 & 0 \\ 0 & -1 \end{bmatrix}$ are respectively the Pauli $X$ and $Z$ matrices; and the full Hamiltonian Operator $H = \sum_{j=1}^{N} H_j$

The first term in Eq. (1) represents the flipping of the qubit from one state to the other, called "tunneling", with amplitude K.

The second term in Eq. (1) represents the energy-difference 2A between the two states. This difference can be the result of external fields or interaction with other qubits. The Pauli Matrix $\sigma_z$ measures the state of the system; e.g. $\sigma_z \begin{bmatrix} 0 \\ 1 \end{bmatrix} = (-1) \begin{bmatrix} 0 \\ 1 \end{bmatrix}$, telling us that the system is in the (-1) state. Thus each qubit $j$ may be visualized as a loop having $n$ discretization points, propagating in imaginary time from 0 to β. [A discretization point at any instant of imaginary time is the instantaneous state of the qubit at that corresponding value of the inverse temperature].

Let $|\psi\rangle = \delta|0\rangle + \gamma|1\rangle$ represent a qubit in terms of Quantum Terminal Graph (QTG) (Srivastava et. al. 2011, 2014) in Figure 4 where $|0\rangle$ is the (+1) state $\begin{bmatrix} 1 \\ 0 \end{bmatrix}$ and $|1\rangle$ is the (-1) state $\begin{bmatrix} 0 \\ 1 \end{bmatrix}$. Notice that $\sigma_x \begin{bmatrix} \delta \\ \gamma \end{bmatrix} = \begin{bmatrix} \gamma \\ \delta \end{bmatrix}$ which flips the probability amplitudes. For example, the (+1) state $|0\rangle$ is flipped to the (-1) state $|1\rangle$, and vice-versa. The Quantum Terminal Graph (QTG) in Fig. 4 (Srivastava et. al. 2011, 2014) is shown as the union of two separate parts such that $a_1 a_1'$ denotes the "ket 0"-part and $b_1 b_1'$ denotes the "ket 1"-part (which form an orthonormal basis for the Hilbert space). Here $\delta$ and $\gamma$ are complex numbers such that $|\delta|^2$ is the probability of qubits of "information / computation data" flow-rate along unit directional (quantum across variable) vector $|0\rangle$; and $|\gamma|^2$ is the probability of "information / computation data" flow-rate along unit directional (quantum across variable) vector $|1\rangle$. Notice that the sum of probabilities of flows along $|0\rangle$ and $|1\rangle$ adds to 1, i.e. $|\delta|^2 + |\gamma|^2 = 1$. Accordingly, the probable qubit states can be said to be normalized to length 1.

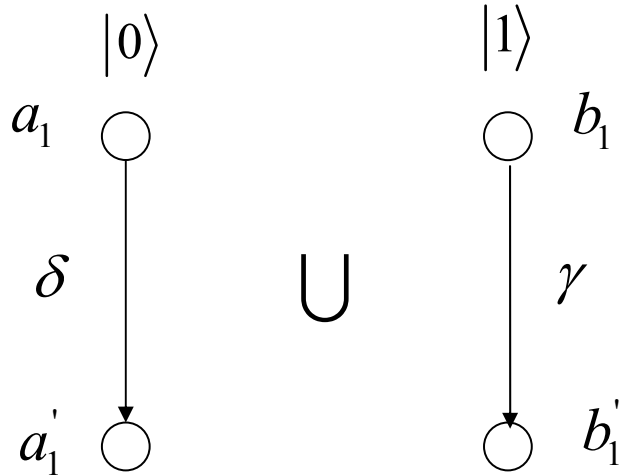

**Figure 4**   Quantum Terminal Graph

Srivastava et. al. (2014) have applied Graph Theoretic Quantum System Modelling (GTQSM) in continuum of protein heterodimer tubulin molecules of self-assembling polymers, viz., microtubules in the brain as a holistic system of interacting components representing hierarchical clustered quantum Hopfield network (hQHN) of networks. The quantum input-output ports of the constituent elemental interaction components (or processes) of tunneling interactions and Coulombic bidirectional interactions are in cascade and parallel interconnections with each other while the classical output ports of all elemental components are interconnected in parallel to accumulate micro-energy functions generated in the system as Hamiltonian (or Lyapunov) energy function. They have presented an insight, otherwise difficult to gain, for complex system of systems represented by clustered quantum Hopfield network (h-QHN), through application of graph theoretic quantum system modelling (GTQSM) construct presented in the next section.

## 4. Generalization of Quantum Hopfield Network Model for n-qudits

Let $|\psi\rangle = \alpha|0\rangle + \beta|1\rangle + \gamma|2\rangle + \delta|3\rangle + ... + \nu|n-1\rangle$ represent a generalized quantum odd prime based qudit (with dimension $d=n$) in comparison to the commonly used even-prime based quantum bit (quits with dimension 2) in terms of Quantum Terminal Graph (QTG) in Figure 5 where $|0\rangle$ is the state $\begin{bmatrix}1\\0\\0\\.\\.\\0\end{bmatrix}$ and $|1\rangle$ is the state $\begin{bmatrix}0\\1\\0\\.\\.\\0\end{bmatrix}$ and $|2\rangle$ is the state $\begin{bmatrix}0\\0\\1\\.\\.\\0\end{bmatrix}$, and $|n\rangle$ is the state $\begin{bmatrix}0\\0\\0\\.\\.\\1\end{bmatrix}$.

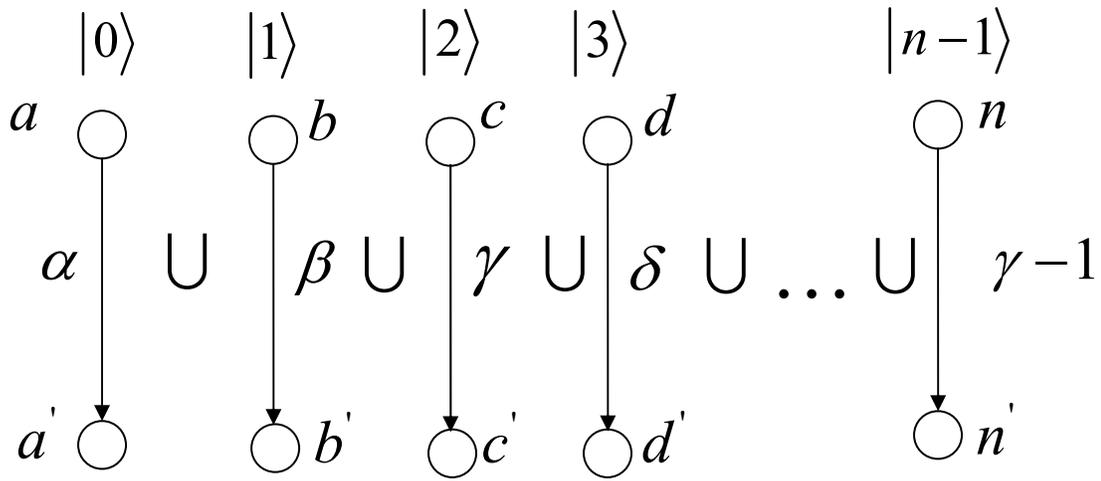

**Figure 5** Generalized Qudit (dimension $d=n$) as union of $n$ separate parts

The generalized quantum terminal graph (QTG) with dimension ($d=n$) is shown in Figure 5 as the union of n separate parts such that $a\,a'$ denote the first "ket-0" part, $b\,b'$ denotes the second "ket-1" part, $c\,c'$ denotes the third "ket-2" part, $d\,d'$ denotes the fourth "ket-3" part and so on, upto $n\,n'$ which denotes the "ket-(n-1)" part (which form an orthonormal basis vector-set for the Hilbert space). Here $\alpha, \beta, \gamma, \delta, (\nu-1)$ are complex numbers such that $|\alpha|^2$ is the probability of qubits of "information / computation data" flow-rate along unit directional (quantum across variable) vector $|0\rangle$; $|\beta|^2$ is the probability of qubits of "information / computation data" flow-rate along unit directional (quantum across variable) vector $|1\rangle$; $|\gamma|^2$ is the probability of qubits of "information / computation data" flow-rate along unit directional (quantum across variable) vector $|2\rangle$; $|\delta|^2$ is the probability of qubits of "information / computation data" flow-rate along unit directional (quantum across variable) vector $|3\rangle$; $|\varepsilon|^2$ is the probability of qubits of

"information / computation data" flow-rate along unit directional (quantum across variable) vector $|4\rangle$ and so on, and $|v-1|^2$ is the probability of "information / computation data" flow-rate along unit directional (quantum across variable) vector $|n-1\rangle$. Notice that the sum of probabilities of flows along $|0\rangle$, $|1\rangle$, $|2\rangle$, ... and $|n-1\rangle$ adds to 1, i.e.

$$|\alpha|^2 + |\beta|^2 + |\gamma|^2 + |\delta|^2 + ... + |v-1|^2 = 1 \qquad (2)$$

Accordingly, the probable qudit states can be said to be normalized to length 1.

Eq. (2) is the equation of an n-dimensional sphere. Thus each qudit $j$ may be visualized as a $d$-dimensional hyper sphere (instead of a loop for qubit which has only 2 states, i.e. $|\alpha|^2 + |\beta|^2 = 1$, a qudit which has only 3 states, i.e. $|\alpha|^2 + |\beta|^2 + |\gamma|^2 = 1$, the equation of a sphere) having $n$ discretization points, propagating in imaginary time from 0 to $\beta$ (this is different from the index $\beta$ for $|1\rangle$). [A discretization point at any instant of imaginary time is the instantaneous state of the qudits at that corresponding value of the inverse temperature].

Figure 6 shows a qubit with $n = 4$ discretization points in a system of $N = 3$ qubits ($d=2$) (Srivastava et. al., 2014).

Figure 7 shows a qudit ($d=3$) with $n = 8$ discretization points in a system of $N = 3$ qudits shown in different colours. Each qubit is now represented by a sphere with 8 discretization points.

Figure 8 shows a qudit ($d=5$) with $n = 32$ discretization points in a system of $N = 5$ qudits. In this solar system model, each 3-dimensional sphere is rotating in a loop / circle and sample connections are shown for discretization points $i = 9$ to 16. Similar connections exist for $i = 1$ to 32 as shown by black lines. If radius of the sphere (planet) is $r_1$ and radius of the orbit is $r_2$ ($r_1$ and $r_2$ are labelled in qudit $j = 3$ for illustration), then

$r_1^2 = |\alpha|^2 + |\beta|^2 + |\gamma|^2$ (equation of a sphere) $p = 3$ degrees of freedom

$r_2^2 = |\delta|^2 + |\varepsilon|^2$ (equation of a loop / circle) $q = 2$ degrees of freedom

Total degrees of freedom $n = p + q = 5$
Discretization Points used = $2^p.2^q = 2^3 \times 2^2 = 8 \times 4 = 32$

We will have $r_1^2 + r_2^2 = 1$ as the system equation for $j$ which is the same as Eq. (2), i.e. $|\alpha|^2 + |\beta|^2 + |\gamma|^2 + |\delta|^2 + |\varepsilon|^2 = 1$ for a 5-dimensional qudit.

This idea can be similarly extended for even higher dimensions of $d$.

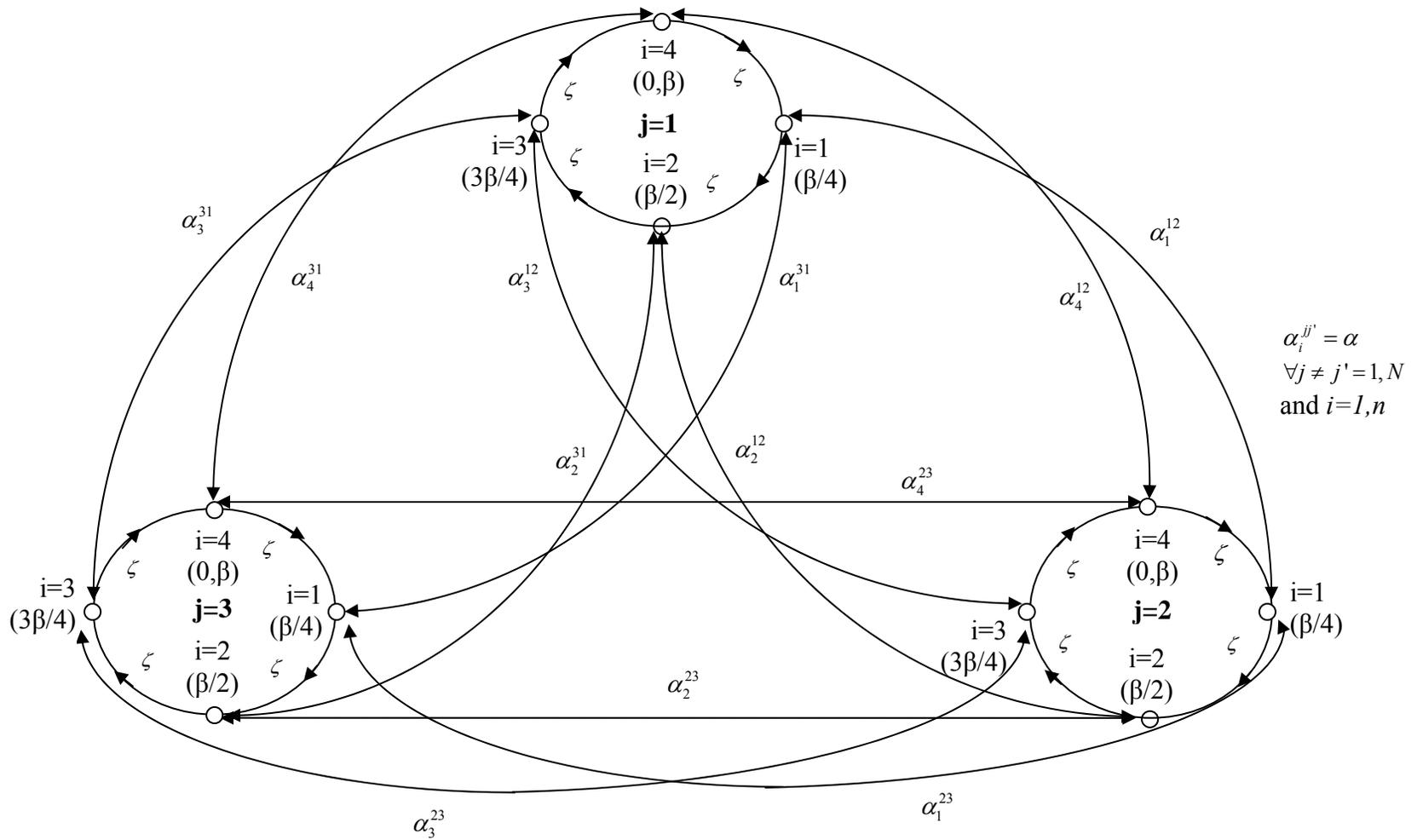

**Figure 6** A qubit with $n = 4$ discretization points in a system of $N = 3$ qubits

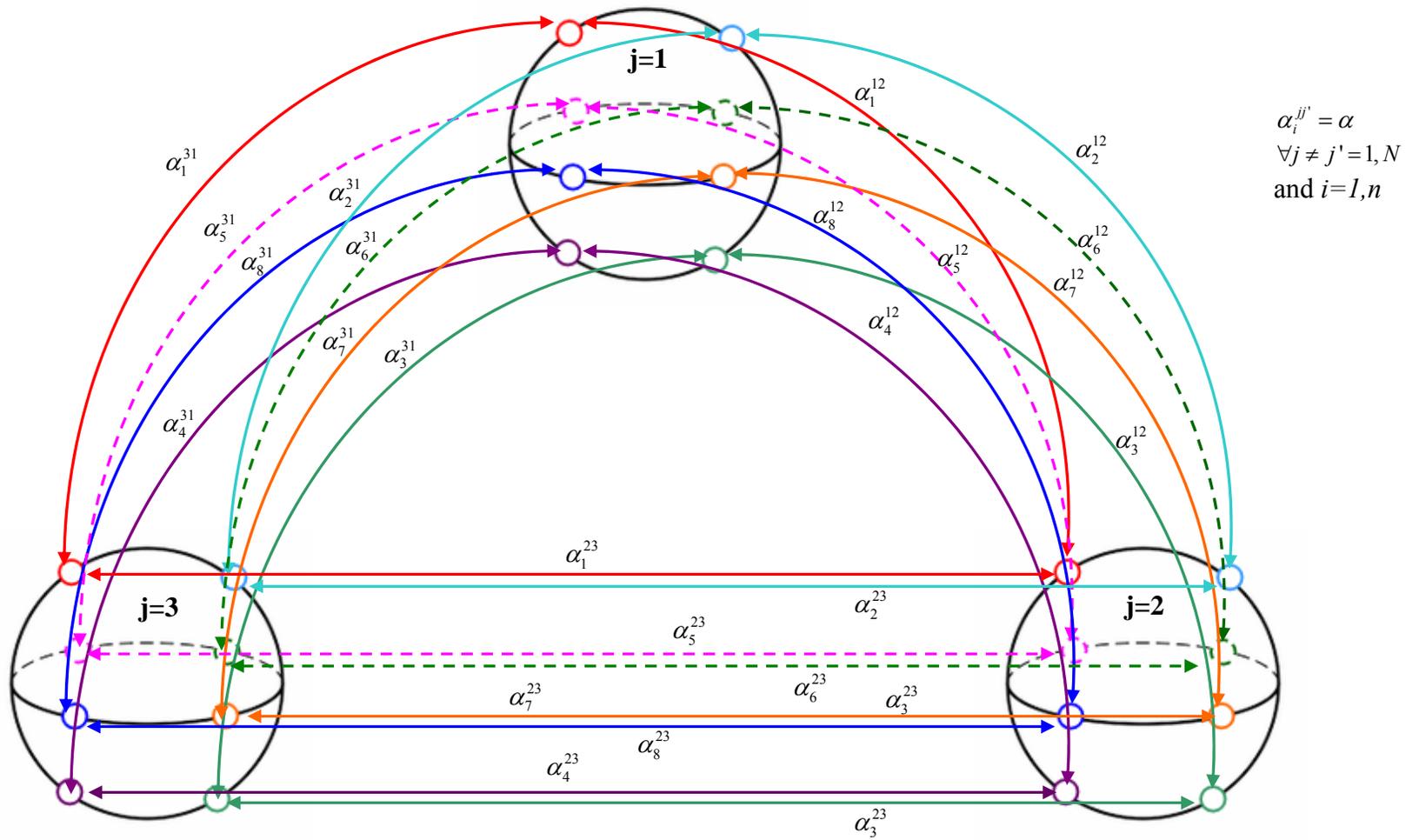

**Figure 7**  A qudit ($d=3$) with $n=8$ discretization points in a system of $N=3$ qudits

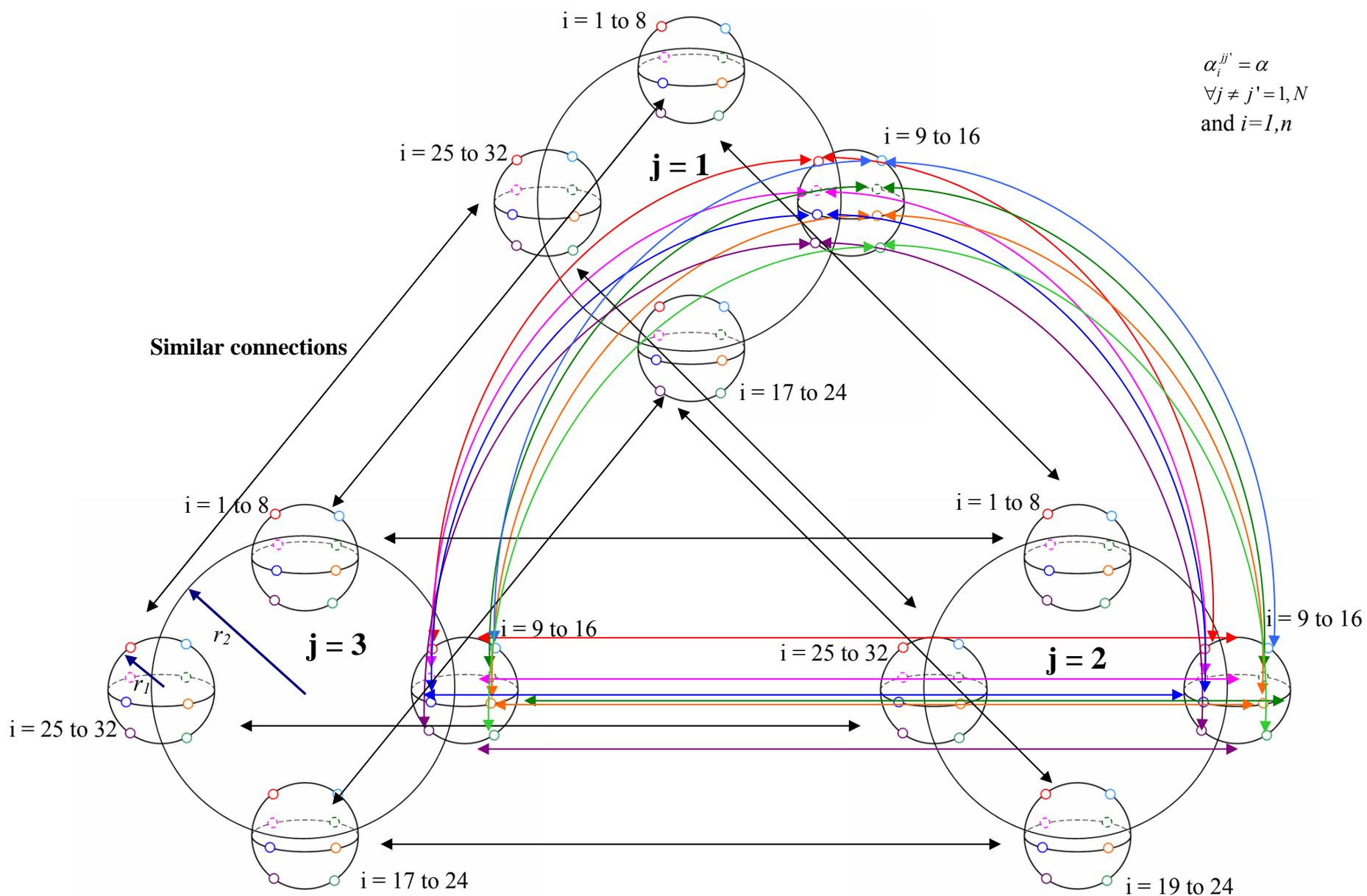

**Figure 8**     A qudit ($d=5$) with $n = 32$ discretization points in a system of $N = 3$ qudits

Sample connections are shown for discretization points $i = 9\ to\ 16$. Similar connections will exist for all $i$ as shown by black lines

## 5. Clustered hierarchical level Model of the Brain

Woolf N. (2006) has proposed five hierarchical levels of quantum entanglement-based neural interaction microtubules spread among 13-23 billion neurons in the human cortex of the estimated 100 billion neurons in brain. We can extends the graph-theoretic modelling framework to these, in general, $l$ levels.

An estimate of $N_l$ and degree of entanglement $\alpha_l$ for $l$ = 1 to 5 by Nancy Woolf (Woolf, 2006) and adapted by Srivastava et. al. (2014) is summarized in Table 1. The cluster approach propounded by Srivastava et. al. (2014) views the quantum Hopfield network as $Nn$ information processing subunits (each with states ±1), corresponding to $N$ qubits of tubulin molecule, which converge to $2^{(Nn)}$ corners of an $(Nn)$-dimensional box.

**Table 1    Hierarchical levels of quantum entanglement**

| Hierarchical Level of Quantum Entanglement | Quantum entanglement among | Estimate of number of qubits $N_l$ by Woolf (2006) as adapted by Srivastava et. al. (2014) | Estimate of $\alpha_l$ by Woolf |
|---|---|---|---|
| 1 | Tubulins of the same microtubule | $N_1 \to 10^4$ tubulins of the same microtubule in any one of $m_1 = 10^{15}$ possible clusters Srivastava et. al. (2014) | likely to be high |
| 2 | Pairs of microtubulins (qubits) in the same neuron | $N_2 \to 10^4$ different microtubules in the same neuron in any one of $m_2 = 10^{15}$ possible clusters Srivastava et. al. (2014) | $\alpha_2 \ll \alpha_1$ |
| 3 | pairs of microtubulins (qubits) belonging to different neurons | $N_3 \to 10^3$ different microtubules belonging to different neurons in same modules in any one of $m_3 = 10^{16}$ possible clusters Srivastava et. al. (2014) | $\alpha_3 \ll \alpha_2$ |
| 4 | highly interconnected cortical areas | $N_4 \to 10^7$ different neurons in any one of $m_4 = 10^{12}$ possible clusters Srivastava et. al. (2014) | $\alpha_4 < \alpha_3$ |
| 5 | entanglement among cortical areas having few or no axonal connections | $N_5 \to 10^{10}$ different neurons in any one of $m_5 = 10^9$ possible clusters (Teleportation invoked by Srivastava et. al. (2014)) | $\alpha_5 < \alpha_4$ |

The first level of neural interaction would include quantum entanglement among approximately $N_1 \rightarrow 10^4$ tubulins of the same microtubule in any one of $m_1 = 10^{15}$ possible clusters. Since this level of neural interaction is between closely interacting particles, the degree or density of entanglement is likely to be high as reflected by the strength $\alpha_1$ of Coulombic bidirectional interaction process between the same discretization points $i$ for each pair of different qubits $j_1$ and $j_1'$ at the first clustered level.

The second level of neural interaction would involve entanglement between pairs of microtubulins (qubits) each drawn from approximately $N_2 \rightarrow 10^4$ different microtubules in the same neuron in any one of $m_2 = 10^{15}$ possible clusters. Microtubule Associated Protein (MAP2) does link neighbouring microtubules together (or to actin filaments) and these linkages dynamically associate and disassociate. This degree of entangelemnt would be somewhat less than that of tubulins in the same microtubule as reflected by strength $\alpha_2 \ll \alpha_1$ of Coulombic bidirectional interaction process at this second clustered level.

The third degree of neural interaction would involve entanglement between pairs of microtubulins (qubits) each drawn from approximately $N_3 \rightarrow 10^3$ different microtubules belonging to different neurons in same modules in any one of $m_3 = 10^{16}$ possible clusters. This degree of entanglement at the third clustered level would be expected to be less than that of tubulins in the different microtubules of the neuron at the second clustered level as reflected by $\alpha_3 \ll \alpha_2$.

The fourth degree of neural interaction would involve entanglement within highly interconnected cortical areas (e.g. cortical areas of the same sensory modality) involving $N_4 \rightarrow 10^7$ different neurons in any one of $m_4 = 10^{12}$ possible clusters. Entanglement can occur over long distances when there is a classical channel of information, for example, axonal connectivity between cortical areas or electromagnetic energy flow from one brain region to another, and a pre-existing entanglement. Qubits performing local operations can even send information to entangled qubits over long distances via classical channels as in quantum teleportation. This degree of entanglement might be expected to be medium to strong as reflected by $\alpha_4 (<\alpha_3)$ at the fourth clustered level.

The fifth level of neural interaction would involve entanglement among cortical areas having few or no axonal connections involving $N_5 \rightarrow 10^{10}$ different neurons in any one of $m_5 = 10^9$ possible clusters. Brain-wide entanglement might occur in a manner predicted by Bell's theorem stating that whenever a quantum measurement is made on one part of a holistic quantum system, this will produce an effect on the other part of the system. This degree of entanglement would be weakest, thereby allowing for greater flexibility in brain response, as reflected by the strengths of Coulombic bidirectional interaction processes at the various clustered levels of hierarchy $\alpha_5 < \alpha_4 < \alpha_3 < \alpha_2 < \alpha_1$.

Quantum entanglement is considered in this section only in neuronal microtubules since non-neuronal cells divide repeatedly and recycle tubulins for mitotic spindle microtubules. On the other hand, neurons don't divide, their microtubules remaining polymerized for the life of the cell, suitable for information encoding, memory and consciousness.

At the first level, within a microtubule, direct quantum information transfer may take place and may be protected through topological or other error-correcting mechanism. At the fourth hierarchical level, where only a classical information channel exists, quantum information transfer and error correction may still take place via quantum teleportation. At the fifth level, quantum error correction may still be possible through weak classical channel and / or collateral learning and training as well as quantum teleportation because of quantum entanglement (Srivastava et. al., 2014).

We model a microtubule as a quantum Hopfield network, an interconnected network consisting of tubulin dimers. Srivastava (2013) and Srivastava et. al. (2014) have represented each tubulin dimer as a qubit. It is essentially quantum mechanical (an Abelian $Z_2$ lattice gauge theory) (Zee, 2010). We can generalize quantum information processing further to quantum dits or qudits that are d-level systems as an extension of qubits that could speed up computing tasks even further. A qubit is then a special case of a qudit with $d = 2$. In comparison to the qubit system, d-dimensional quantum states will be more efficient in quantum applications. In our ongoing work, we are considering tubulin dimers as higher order qubits called qudits, which are quantum systems of dimensionality d (an odd prime). So, the values of d, such as 3 and 5, this model is quantum field theoretic (a non-abelian lattice gauge theory). For larger values of d, our model may be string theoretic. Graph theory has helped us to address the problem of modelling continuum, which is essentially a field problem by converting it into a circuit problem by discretization, which is an approximation made by us using graph theory. Graph theory allows for multi-terminal representations, aggregation and disaggregation, thereby allowing us to model the brain at any of the five hierarchical levels. A system schematic for hierarchical clustered quantum Hopfield networks appears in Figure 9 (Srivastava 2013, Srivastava et. al., 2014).

The Hamiltonian (or Lyapunov) Energy Function $H_{hQHN}$ for the simplest possible Quantum Hopfield Network, considering $N_l$ qubits (at clustered hierarchy level $l$) each with $n$ discretization points, propagating in imaginary time (inverse temperature) over 0 to $\beta$, (where each pair of qubits interacts only at equal imaginary discretized time $i$), is given by (Srivastava 2013, Srivastava et. al., 2014) :

$$H_{hQHN} = \left\{ H_I = \sum_{j_0=1}^{N_0} \sum_{i=1}^{n} H_{i+1}^{j_0} \right\} + \left\{ H_{II} = \sum_{j=1}^{N_l} \sum_{i=1}^{n} H_i^{j_l, j_l'} \right\}$$

$$= \left\{ -\frac{1}{\beta} \zeta \left( \sum_{j_0=1}^{N_0} \sum_{i=1}^{n} S_i^{j_0} S_{i+1}^{j_0} \right) \right\} + \left\{ -\frac{1}{\beta} \sum_{l=1}^{5} \left[ \alpha_l \left( \sum_{j \neq j_l=1}^{N_l} \sum_{i=1}^{n} S_i^{j_l} S_i^{j_l'} \right) \right] \right\} \quad (3)$$

such that $N_0 = \bigcup_{l=1,5} N_l \leq \sum_{l=1}^{5} N_l$

The first term in Eq. (3) is tunnelling energy and the second is Coulombic energy.

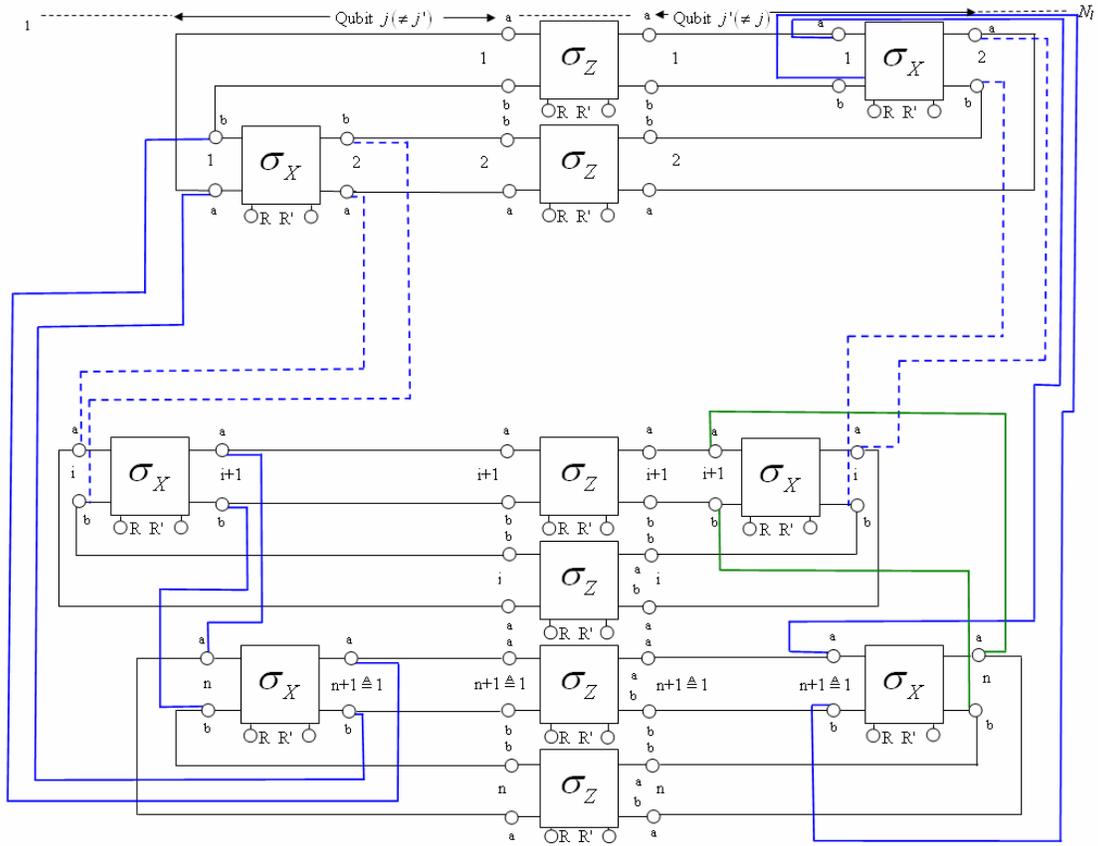

**Figure 9**   Hierarchical Clustered Quantum Hopfield Network (hQHN) – System Schematic

## 6. Contextuality

In a recent paper in Nature, Howard et. al. (2014) show that contextuality, an abstract generalization of nonlocality is a necessary condition for magic state distillation. They make an abstract graph called an exclusivity graph to obtain their results. Qudits of odd prime dimensionality are naturally differentiated from qubits in their analysis. In exclusivity graph in Figure 10(a), each of the 30 vertices corresponds to a two-qubit stabilizer state; connected vertices correspond to orthogonal states. A maximum independent set (representing mutually non-orthogonal states) of size 8 is highlighted in red. They have proved a remarkable equivalence between the onset of contextuality and the possibility of universal quantum computation via 'magic state' distillation, which is the leading model for experimentally realizing a fault-tolerant quantum computer. This is a conceptually satisfying link, because contextuality, which precludes a simple 'hidden variable' model of quantum mechanics, provides one of the fundamental characterizations of uniquely quantum phenomena. Furthermore, this connection suggests a unifying paradigm for the resources of quantum information: the non-locality of quantum theory is a particular kind of contextuality, and non-locality is already known to be a critical resource for achieving advantages with quantum communication.

We are working on simulating a functional model of the brain based on quantum Hopfield networks with neurons selected from the various regions of the brain, based on functional layer diagram. Such models should be able to handle association and contextuality both as they will have neurons from portions of the brain tasked with association (association cortex) as well as contextuality (pre-frontal cortex). Table 2 gives a summary of the macrocolumn functional units mentioning number of columns, which is believed to be the fundamental computational unit of the cortex, and an estimate of the neurons in various lobes of the brain.

**Table 2**
**(a)    Macrocolumn functional units of the cortex**

| Functional Unit | Columns | Billion Neurons |
|---|---|---|
| Visual Cortex | 232 million columns | 68.3 |
| Auditory Cortex | 83 million columns | 24.0 |
| Somatosensory cortex | 21 million columns | 6.0 |
| Association cortex | 5 million columns | 1.4 |
| Motor cortex | 1 million columns | 0.3 |
| TOTAL | 342 million columns | 100 billion |

**(b) An estimate of the neurons in various lobes of the brain**

| | |
|---|---|
| Frontal Lobe | 35 billion |
| Motor cortex (precentral gyms) | 6 billion |
| Temporal Lobe | 19 billion |
| Parietal Lobe | 16 billion |
| Occipital Lobe | 14 billion |
| TOTAL | 90 billion |

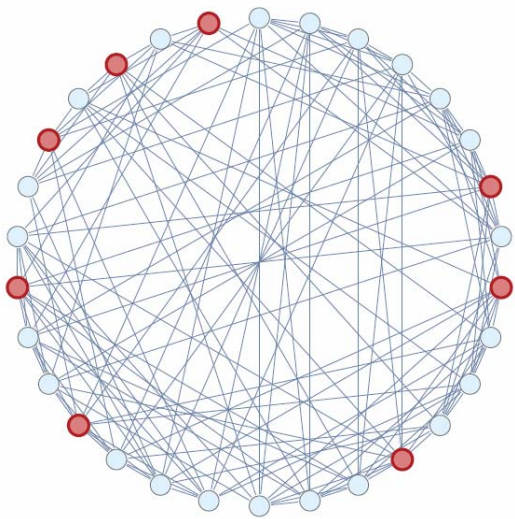

(a) Exclusivity graph proposed by Howard et. al. applied to two qubits

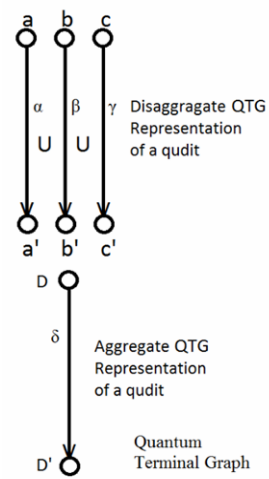

(b) QTG Representation

**Figure 10**

Based on insights given from Srivastava (2013) and Srivastava et. al. (2014), it is proposed to derive exclusivity graph in Figure 10(b) by representing Quantum Terminal Graphs (Srivastava 2013, Srivastava et. al. 2014) as composite Quantum Terminal Graphs each represented by a single composite quantum terminal graph which may be further reduced to a representation in terms of a single composite quantum vertex or node as in Figure 10(b) by rendering the composite datum node as a floating unspecified datum node. This representation was also presented by a team from DEI before Department of Science and Technology, Govt. of India's Cognitive Science Research Initiative (CSRI) Review Committee in March 2015 and revised project proposal prepared in April 2015.

We believe that our graph theoretic approach based on physical systems theory, will give a simpler derivation of their results potentially allowing us to demonstrate that contextuality is not only necessary but also sufficient for magic state distillation. We are attempting to address the important question of how does the brain overcome decoherence using topological, fault tolerant computing and magic state distillation.

## 7. Power Laws

To gain a picture of emergent patterns of brain activity, investigators need new sensing devices that can record from assemblages of thousands of neurons. Nanotechnology, with novel materials that sometimes measure less than the dimensions of individual molecules, may assist in making large-scale recordings. Prototype arrays have been built that incorporate more than 100,000 electrodes on a silicon base; such devices could record the electrical activity of tens of thousands of neurons in the retina.

Electrodes are only one way to track the activity of neurons. Methods that move beyond electrical sensors are making their way into the lab. Biologists, borrowing from technologies developed by physicists, chemists and geneticists, are beginning to visualize living neurons in awake animals going about their daily paces using electro-encephalography (EEG), magento-encephalography (MEG) and functional Magentic Resonance Imaging (fMRI) and further analysis by Fourier transform methods.

An interesting conclusion we can draw from a course on noise is that the $1/\rho$ behaviour of EEG and MEG (i.e. "one over of" power-spectrum) is a special noise (also called 'pink' noise) is the golden means between the disorder with high information content (white noise with $1/\rho^0$ flat (constant power spectrum) and the predictability with low information content (brown noise with "one over $\rho^2$ power spectrum"). The cerebral cortex with its most complex architecture generates the most complex noise known to physics (Buzsaki, 2006).

The anatomical-functional organization of the cerebral cortex should have consequences and limitations on cognitive behaviours as well. A well-known psychological law that comes to mind in connection with the 1/$f$ nature of the cortical EEG / MEG is that of Weber and Fechnel : the magnitude of a subjective sensation (a cognitive unit) increases proportionally to the logarithm of the stimulus intensity (a physical unit).

This collective behaviour of neurons, summed up crudely as the mean field (EEG / MEG) is a blend of rhythms. Neuronal network in the mammalian cortex generates several distinct oscillatory bands, covering frequencies from <0.05 hertz to >500 hertz.

Brain evolution opted for a complex wiring pattern in the mammalian cortex. The resulting $1/f^{\alpha}$ temporal statistic of the mean field are the hallmark of the most complex dynamics and imply an inherently labile, self-organized state.

Unlike so-called normal distribution (such as for human brain weight (e.g. 1.35 kg or 3 lbs.), in scale-free systems (governed by power law), things are different. There is no typical example in a scale-free system. A power law implies for the case of fracture, that if we plot the numbers of different size of pieces on a log-log scale, we will get an oblique line.

In fractal geometry, fractals are usually defined in statistical or qualitative terms, loosely indicating anything that "looks like itself". The scale invariance of fractals implies that knowledge of the properties of a model system at any scale can be used to predict the structure of the real system at larger or smaller scale.

## 8. Conclusions

Price and Barrell (2012) show how a science of human experience can be developed through a strategy by integrating experiential paradigms with methods from the natural sciences. A science of human experience would largely be about human meanings and also would constitute a science of consciousness. Psychophysics includes a rudimentary form of introspection that is critical for developing a science of human meanings and consciousness. The qualitative aspects of human experience serve as a foundation for developing hypotheses to be used in experiments that have quantitative methods. This model serves as a useful example for correlating first person and third person approaches for explaining a conscious experience.

Interesting conclusions can be drawn from a study of $1/f^{\alpha}$ behaviour of brain-scans of humans, particularly during Eastern Meditational Practice. The cerebral cortex with its most complicated architecture generates the most complex noise known to Physics. One over $f$ ($\alpha \equiv 1$) power spectrum (also called "pink noise") is a golden mean between the white noise ($\alpha \equiv 0$, i.e. constant power spectrum) representing disorder with high information content, and brown noise ($\alpha \equiv 2$) representing predictability with low information content. The anatomical functional organization of the cerebral cortex brings to mind a well-known psycho-physical law credited to Weber and Fechner. Accordingly, the magnitude of a subjective sensation such as meditational consciousness in cognitive and meta-cognitive domains increases proportionally to the logarithm of the stimulus intensity, "one over $f$ (i.e. rhythmic frequency)", a physical unit. Notably, scale-free systems or fractals are governed by Power Laws (Satsangi and Sahni, 2015).

The formulation of power law of meditational consciousness requires invoking a family of models based on Omni quantum theory and physical system theory (including fuzzy analytical hierarchy process), and requisite integration of first person inner experiences of meditationists as co-investigators with the third person scientific methodology of observing, reporting, understanding and hypothesis-testing (Satsangi and Sahni, 2015).